# ANISOTROPIC SPHERES WITH BAROTROPIC EQUATION OF STATE IN BIMETRIC THEORY OF RELATIVITY

**Naveen Bijalwan[1]**

Recently, Khadekar (2007) presented the solutions with uniform energy density for anisotropic spheres in bimetric theory. We present here a general analytic solution to the field equations in bimetric theory for anisotropic fluids for a general barotropic equation of state by representing equations in terms for effective radial pressure $p_r$. We list and discuss some old and new solutions which fall in this category.

## 1. Introduction

The study of static anisotropic spheres is important in relativistic astrophysics (Bowers and Liang 1974). In general relativity Maharaj and Maartens (1989) have studied a class of solutions for static anisotropic spheres with uniform energy-density. Recently, Maharaj(2006) presented the solutions that satisfy a barotropic equation of state linearly for anisotropic spheres. Rosen (1979, 1980) proposed a new theory of gravitation called the bimetric general relativity theory. Several authors have studied various aspects of biometric theory of gravitation with matter (Reddy and Venkateswarlu 1989; Reddy and Venkateswara Rao 2000; Mohanty and Sahoo 2002; Reddy 2003, Khadekar 2007).

In this paper we present a class of solutions for anisotropic spheres in bimetric relativity theory with general barotropic equation of state in terms of effective radial pressure imitating Bijalwan (2011) ansatz.

## 2. Field Equations

Let us take the following spherically symmetric metric to describe the space-time of an anisotropic fluid

$$ds^2 = -e^{\lambda} dr^2 - r^2(d\theta^2 + \sin^2\theta d\phi^2) + e^{\upsilon} dt^2 \qquad (2.1)$$

with $\upsilon$ and $\lambda$ are functions only of $r$.

The energy-momentum tensor is of the form

$$T^{ij} = \mu u^i u^j + p h^i h^j + \pi^{ij} \qquad (2.2)$$

where $\mu$ is the energy-density, $p$ is the isotropic pressure, $u^i = e^{-\upsilon/2}\delta^i_t$ is the four-velocity vector of the fluid, $h^{ij} = g^{ij} + u^i u^j$ is the projection tensor and $\pi^{ij}$ is the anisotropic pressure (stress) tensor.

The field equations with respect to the metric (2.1) reduce to (eq. 16-18) Khadekar et al (2007)

$$-\frac{\upsilon'}{r}e^{-\lambda} + \frac{(1-e^{-\lambda})}{r^2} = -p_r, \qquad (2.3)$$

[1]FreeLancer, c/o Sh. Rajkumar Bijalwan, Nirmal Baag, Part A, Pashulock, Virbhadra, Rishikesh, Dehradun-249202 (Uttarakhand), India. ahcirpma@rediffmail.com

$$\left[\frac{v''}{2} - \frac{\lambda' v'}{4} + \frac{v'^2}{4} + \frac{v' - \lambda'}{2r}\right] e^{-\lambda} = p_\perp, \tag{2.4}$$

$$\frac{\lambda'}{r} e^{-\lambda} + \frac{(1 - e^{-\lambda})}{r^2} = \mu, \tag{2.5}$$

where $\mu$ is the effective density and, $p_r$ and $p_\perp$ are pressures respectively. Primes denote differentiation with respect to $r$.

Let us consider the barotropic equation of state $\mu = g(p_r)$. \hfill (2.6a)

On subtracting (2.3) from (2.5) gives

$$\left(\frac{v' + \lambda'}{r}\right) e^{-\lambda} = (\mu + p_r) \tag{2.7a}$$

$$\left(\frac{v' + \lambda'}{r}\right) e^{-\lambda} = (g + p_r) \tag{2.7b}$$

Now, in order to solve (2.7b), let us further assume that metrics ($e^\lambda$ and $e^v$), and pressure $p_\perp$ are arbitrary functions of pressure $p_r(\omega)$ such that $\omega$ is some function of $r$ i.e.

$$e^{-\lambda} = s(p_r(\omega)), \ e^v = h(p_r(\omega)), \mu = g(p_r(\omega)) \tag{2.6b}$$

Substituting (2.6b) in (2.7b) leads to

$$\frac{(\bar{v} + \bar{\lambda})}{(\mu + p)} e^{-\lambda} \frac{dp_r}{dr} = r \tag{2.8}$$

where overhead dash denotes derivative w.r.t. $p_r$ or $\omega$.

(2.8) yields

$$p_r = f(c_1 + c_2 r^2) = f(\omega)$$

i.e. function of '$c_1 + c_2 r^2$', where $c_1$ and $c_2 (\neq 0)$ are arbitrary constants, such that

$$r = \sqrt{\frac{\omega - c_1}{c_2}}, \ (\omega - c_1) c_2 > 0 \tag{2.9}$$

Assuming $f$ is invertible and $f^{-1}$ is inverse of $f$ then $f^{-1}(p_r) = \omega$.

Effective density tangential pressures can be expressed using (2.6) in (2.3), (2.4) and (2.5) as

$$\mu = c_2 \left(\frac{\bar{h}}{h} s - \bar{s}\right) - p_r(\omega) \tag{2.10}$$

$$p_r = c_2 \left(2s \frac{\bar{h}}{h} - \frac{(1 - s)}{(\omega - c_1)}\right) \tag{2.11a}$$

or $\ h = \exp\left(\int \frac{1}{s}\left(\frac{(1 - s)}{(\omega - c_1)} + \frac{p_r}{c_2}\right)\right) \tag{2.11b}$

$$\left[2\frac{\bar{h}}{h}+(\omega-c_1)\overline{\left(\frac{\bar{h}}{h}\right)}+(\omega-c_1)\left(\frac{\bar{s}}{s}\right)\left(\frac{\bar{h}}{h}\right)+(\omega-c_1)\left(\frac{\bar{h}}{h}\right)^2+\frac{\bar{s}}{s}\right]s=\frac{p_\perp}{c_2} \qquad (2.12)$$

Equation (2.12) is first order linear differential equation for $s$ i.e.
$$\bar{s}+Zs=Z_1 \qquad (2.13)$$
where

$$Z=\frac{[2\alpha+f_1\bar{\alpha}+f_1\alpha^2]}{[f_1\alpha+1]}, Z_1=\frac{\left[\frac{p_\perp}{c_2}\right]}{[f_1\alpha+1]}, \text{ such that } \alpha=\frac{\bar{h}}{h} \text{ and } f_1(p)=\omega-c_1$$

Solving (2.13) leads to
$$s=e^{-\int zd\omega}\int e^{\int zd\omega}Z_1 d\omega \qquad (2.14)$$

**TABLE 1: METRICS FOR ANISOTROPIC FLUIDS**

| $e^\nu$<br>[name [ref.]] | $e^{-\lambda}(=e^{-\int zd\omega}\int e^{\int zd\omega}Z_1 d\omega)$ (neutral or charged)<br>$f_1(p)=\omega-c_1$, $p_\perp=p_\perp(\omega)$ |
|---|---|
| $=c_4\exp\left(\left(1-\frac{(\omega-c_1)}{c_2 R^2}\right)^{n+1}\right)\left(1-\frac{2M(\omega-c_1)}{c_2 R^3}\right)^{-1/2}$<br><br>$=c_4\exp\left(\left(1-\frac{r^2}{R^2}\right)^{n+1}\right)\left(1-\frac{2Mr^2}{R^3}\right)^{-1/2}$<br><br>*[Khadekar (2007)]* | $Z=\frac{[2\alpha+f_1\bar{\alpha}+f_1\alpha^2]}{[f_1\alpha+1]}, Z_1=\frac{\left[\frac{p_\perp}{c_2}\right]}{[f_1\alpha+1]}$<br><br>$\alpha=-\frac{(n+1)}{c_2 R^2}+\frac{M}{c_2 R^3\left(1-\frac{2M(\omega-c_1)}{c_2 R^3}\right)}$ |

## 3. Conclusions
We have studied the structure of the sources produced by field equations in bimetric theory with an anisotropic matter distribution for barotropic equation of state. Classes of physically acceptable analytic solutions for different reasonable choices of the radial pressure can be obtained.